\input{aipcheck}

\documentclass[
    ,final            
  ]
  {aipproc}

\layoutstyle{8x11single}

\begin{document}

\title{Do WMAP data favor neutrino
mass and a coupling between Cold Dark Matter and Dark Energy$\, $?}



\author{S. A. Bonometto}{
  address={Department of Physics G.~Occhialini -- Milano -- Bicocca
University, Piazza della Scienza 3, 20126 Milano}
,altaddress={I.N.F.N., Sezione di Milano} }

\author{G. La Vacca}{
  address={Department of Physics G.~Occhialini -- Milano -- Bicocca
University, Piazza della Scienza 3, 20126 Milano}
,altaddress={I.N.F.N., Sezione di Milano} 
}

\author{J. R. Kristiansen}{
  address={Institute of Theoretical
Astrophysics, University of Oslo, Box 1029, 0315 Oslo, Norway}
}

\author{R. Mainini}{
  address={Institute of Theoretical
Astrophysics, University of Oslo, Box 1029, 0315 Oslo, Norway}
}

\author{L. P. L. Colombo}{
  address={Department of Physics \& Astronomy,
University of Southern California, Los Angeles, CA 90089-0484}
}

\begin{abstract}
We fit WMAP5 and related data by allowing for a CDM--DE coupling and
non--zero neutrino masses, simultaneously. We find a significant
correlation between these parameters, so that simultaneous higher
coupling and $\nu$--masses are allowed. Furthermore, models with a
significant coupling and $\nu$--mass are statistically favoured in
respect to a cosmology with no coupling and negligible neutrino mass
(our best fits are: $C \sim 1/2m_p$, $ m_\nu \sim 0.12\, $eV per
flavor). We use a standard Monte Carlo Markov Chain approach, by
assuming DE to be a scalar field self--interacting through
Ratra--Peebles or SUGRA potentials.
\end{abstract}

\maketitle


\section{Why DE and CDM should be coupled}

One of the main puzzles of cosmology is why a model as $\Lambda$CDM,
implying so many conceptual problems, apparently fits all linear data
\cite{bib1,bib2,bib3} in such unrivalled fashion.

It is then important that the fine tuning paradox of $\Lambda$CDM is
eased in cosmologies where DE is a self--interacting scalar field
$\phi$ (dDE cosmologies), with no likelihood downgrade \cite{bib6}.
The coincidence paradox is also eased in coupled DE (cDE) cosmologies,
{\it i.e.}, if an energy flow from CDM to dDE occurs
\cite{coupling}. CDM--DE coupling, however, cuts the model
likelihood, when we approach a coupling intensity significantly
attenuating the coincidence paradox.

The physical cosmology could however include a further ingredient,
able to compensate coupling distorsions. Here we show that a possible
option is neutrino mass. In fact, when we assume CDM--DE coupling
or a significant $\nu$ mass, we cause opposite spectral shifts
\cite{lavacca}.

It is then natural to try to compensate them; if we do so, however,
the residual tiny distorsions tend to favor coupling and $\nu$ mass,
in respect to dDE or $\Lambda$CDM.

We illustrate this fact by considering the self--interaction potentials
\begin{equation}
V(\phi) = \Lambda^{\alpha+4}/\phi^\alpha 
\end{equation}
or
\begin{equation}
V(\phi) = (\Lambda^{\alpha+4}/\phi^\alpha) \exp(4\pi\, \phi^2/m_p^2),
\end{equation}
(RP \cite{RP88} and SUGRA \cite{SUGRA}, respectively; $m_p:$ the
Planck mass), admitting tracker solutions. Uncoupled RP (SUGRA) yields
a slowly (fastly) varying $w(a)$ state parameter. Coupling is however
an essential feature and modifies these behaviors, mostly lowering
$w(a)$ at low $z$, and boosting it up to +1, for $z >\sim 10$.

For any choice of $\Lambda$ and $\alpha$ these cosmologies have a
precise DE density parameter $\Omega_{de}$. Here we take $\Lambda$ and
$\Omega_{de}$ as free parameters in flat cosmologies; the related
$\alpha$ value then follows.

In these scenarios, DE energy density and pressure read
\begin{equation}
\rho = \rho_k + V(\phi)~,~~~ p = \rho_k - V(\phi)~,
\end{equation}
{\rm with}~~
\begin{equation}
\rho_k = \dot \phi^2/2a^2~;
\label{rhop}
\end{equation}
dots indicating differentiation in respect to $\tau$, the background
metrics being
\begin{equation}
ds^2 = a^2(\tau) \left[ d\tau^2 - d\lambda^2 \right]
\label{metric}
\end{equation}
with
\begin{equation}
d\lambda^2 = dr^2 + r^2(d\theta^2 + cos^2 \theta\, d\phi^2)~.
\end{equation}
Until $\rho_k \gg V$, therefore, $w(a)=p/\rho$ approaches +1 and DE
density would rapidy dilute ($\rho \propto a^{-6}$), unless a feeding
from CDM occurs. The gradual increase of $\phi$ leads it to approach
$m_p$, so that a $V \gg \rho_k$ regime is attained. The state
parameter approaches then --1 and DE induces cosmic acceleration.

DE cannot couple to baryons, because of the equivalence principle
(see, {\it e.g.}  \cite{darmour1}). Constraints to CDM--DE
interactions, however, can only derive from cosmological data.
\begin{figure}
\includegraphics[height=6.8cm,width=8.cm]{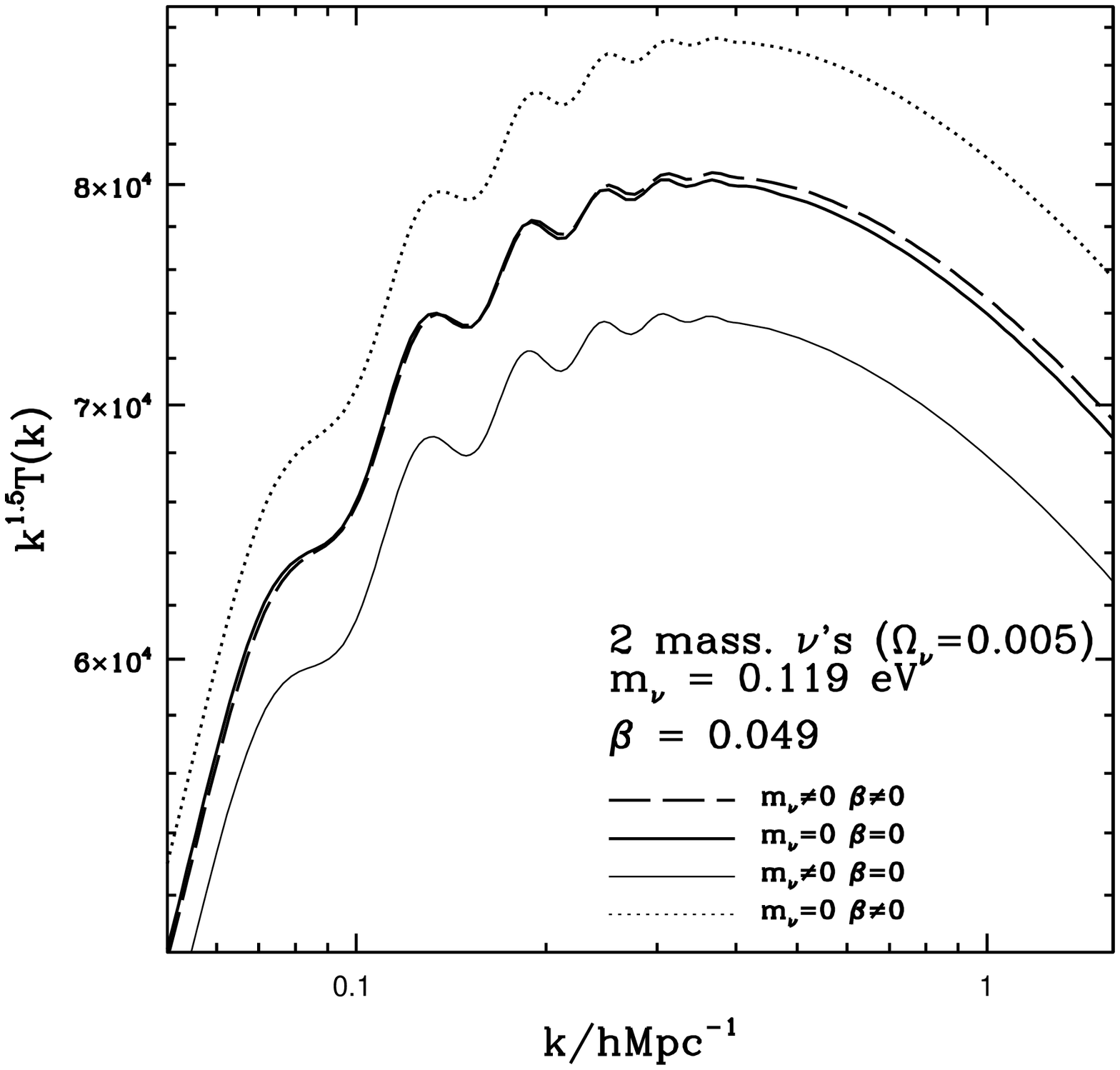}
\includegraphics[height=8.5cm,width=8.cm]{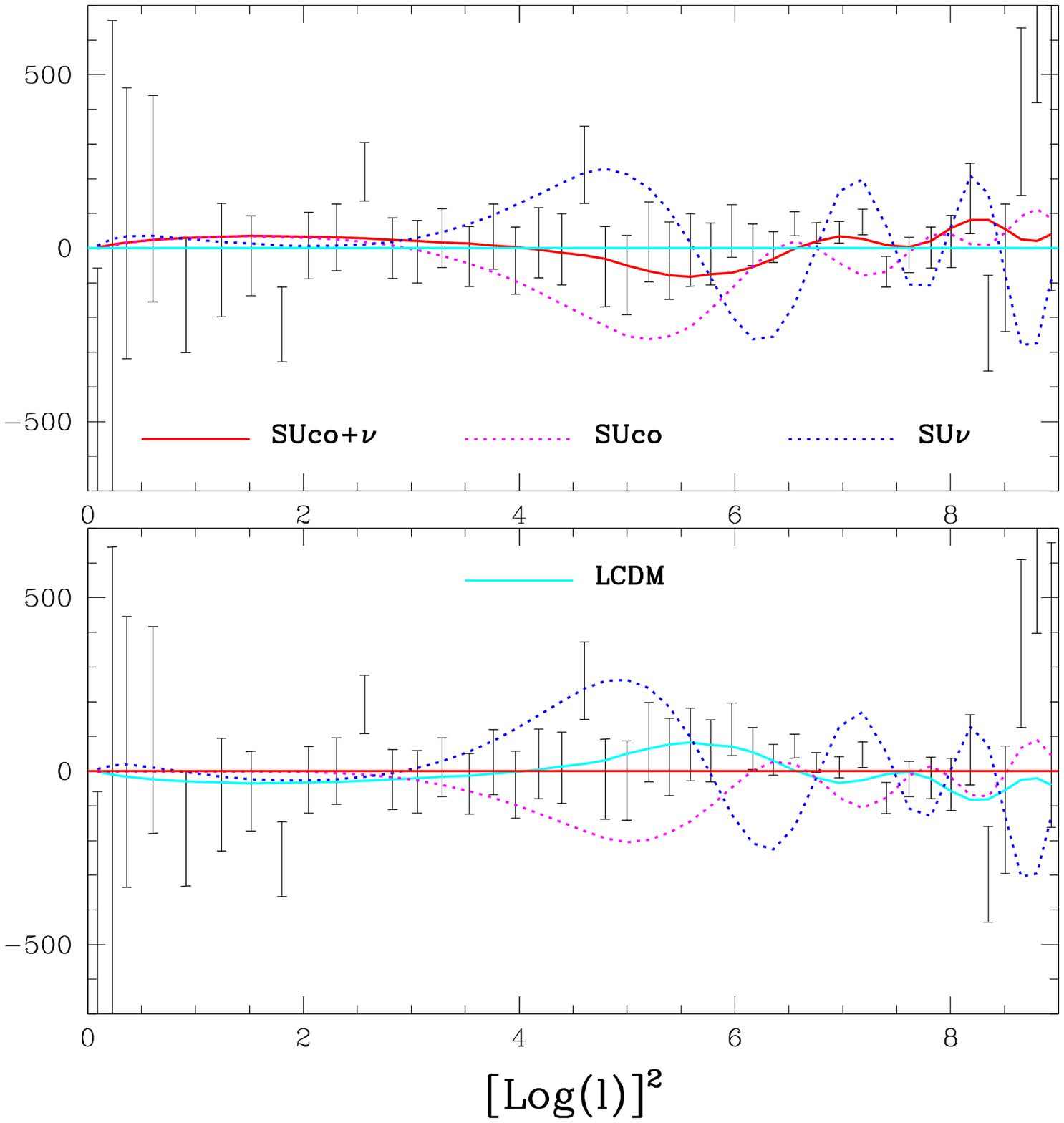}
\caption{{\bf Left panel:} Transfer functions (multiplied by $k^{1.5}$,
for graphic aims) in example cosmologies with/without coupling and
with/without 2 massive $\nu$'s (00/CM models). {\bf Right panel:}
Binned anisotropy spectral data, normalized to the best--fit
$\Lambda$CDM model (upper frame) or to the best fit SUGRA cDE model
including $\nu$--masses (lower frame).  The distorsions arising when
coupling or $\nu$--mass are separately considered are also shown.}
\label{fig1a}
\end{figure}
The simplest coupling is a linear one, formally obtainable through a
conformal transformation of Brans--Dicke theory (see, {\it e.g.},
\cite{brans}). The coupling intensity must however be adequate
to transfer from CDM to DE the energy needed to beat its spontaneous
dilution $\propto a^{-6}$. This prescribes a coupling scale $C^{-1}
\sim m_p$, while the energy drain from CDM lets its density decline
(slightly) more rapidly than $a^{-3}$. The whole scenario between
recombination and the birth of non linearities is then modified, and
it comes as no surprise that we expect significant changes both
in the matter density fluctuation spectra $P(k)$ and in the
observed angular anisotropy spectrum $C_l$.
\begin{figure}
\includegraphics[height=7.2cm,angle=0]{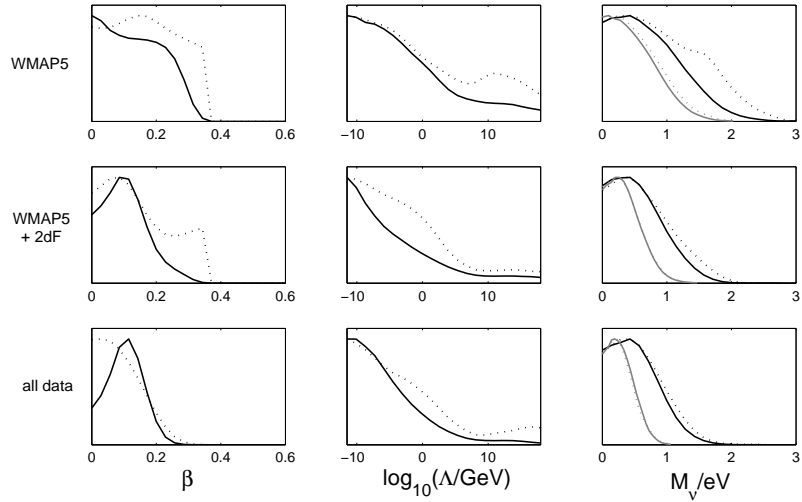}
\caption{Marginalized (solid line) and average (dotted line)
likelihood of cosmological parameters in SUGRA models.  Notice that
the $\beta$ signal appears when high-- and low--$z$ data are put
together, and is strengthened by SNIa data.  As a matter of fact,
coupling allows to lower the ``tension'' between $\Omega_c$ and
fluctuation amplitude detected from CMB and deep sample data. For
$M_{\nu}$ we also show the corresponding likelihood distributions
obtained in the case of a standard $\Lambda$CDM+$M_\nu$ model (gray
lines).}
\label{mnuall}
\end{figure}
\begin{figure}
\includegraphics[height=7.2cm,angle=0]{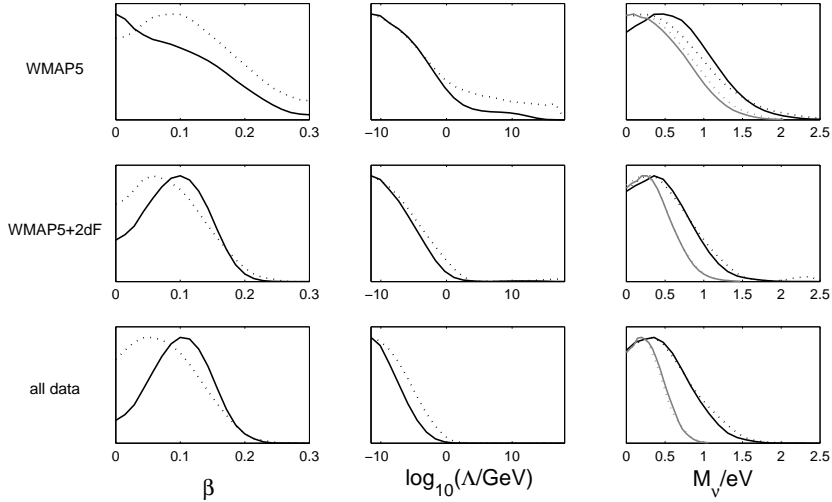}
\caption{As previous Figure, in RP models.}
\label{Rmnuall}
\end{figure}
\begin{figure}
\includegraphics[height=6.5cm,width=7.cm,angle=0]{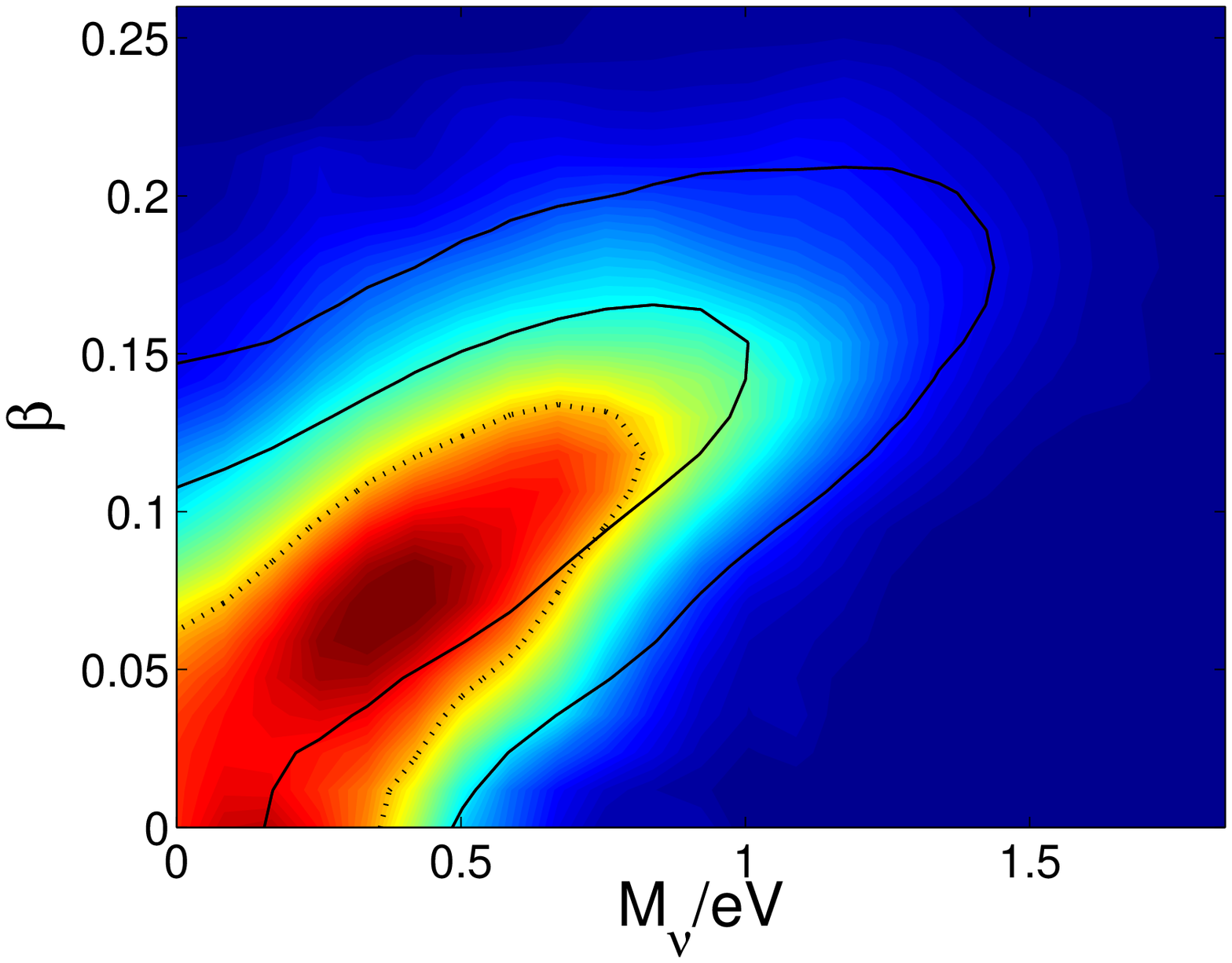}
\includegraphics[height=6.5cm,width=7.cm,angle=0]{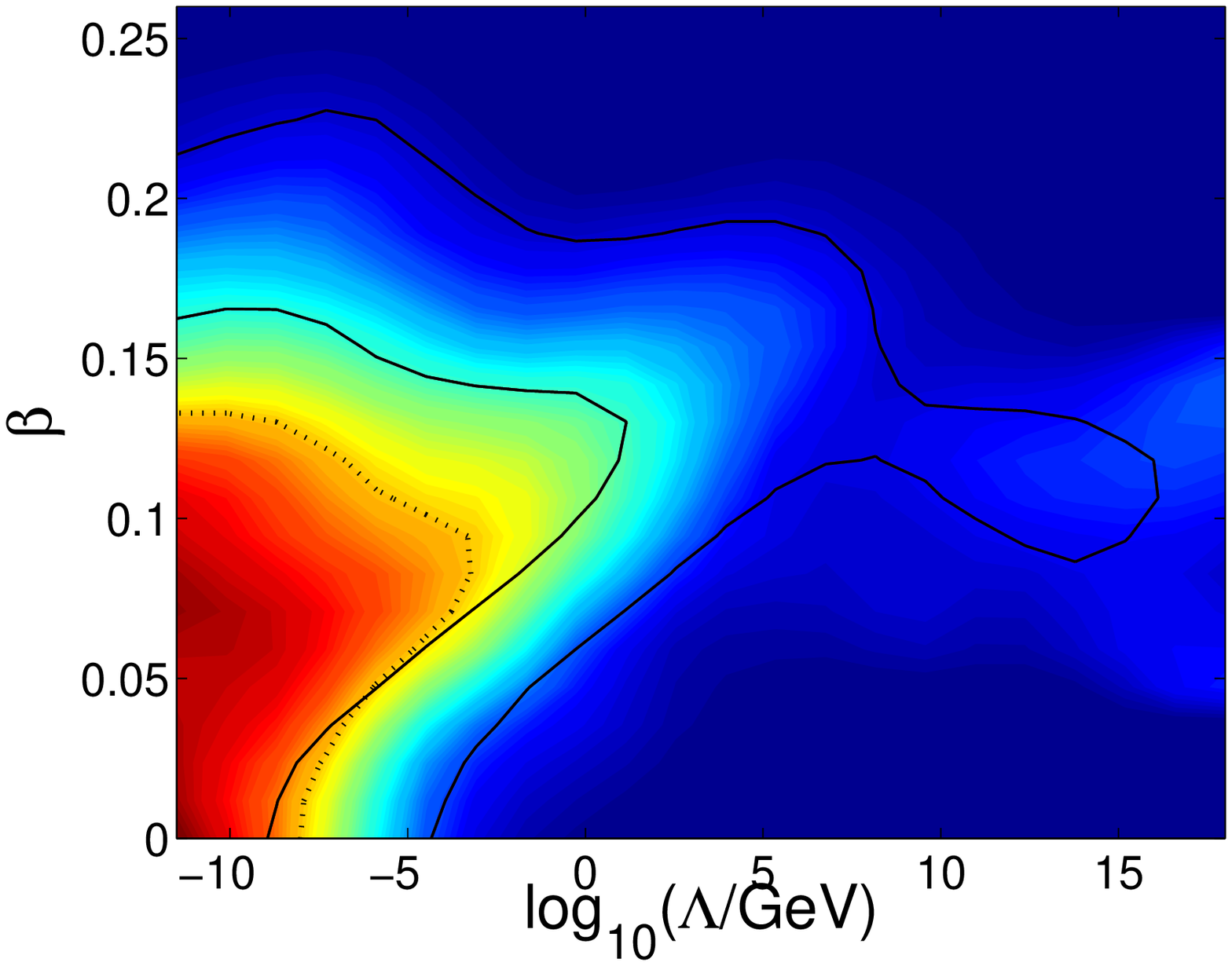}
\caption{Two parameter contours for the SUGRA model. Solid lines are
1-- and 2--$\sigma$ limits for marginalized likelihood.  Colors refer
to average likelihood, and the 50\% likelihood contour from the
average likelihood is indicated by the dotted line.}
\label{2DS}
\end{figure}
\begin{figure}
\includegraphics[height=6.5cm,width=7.cm,angle=0]{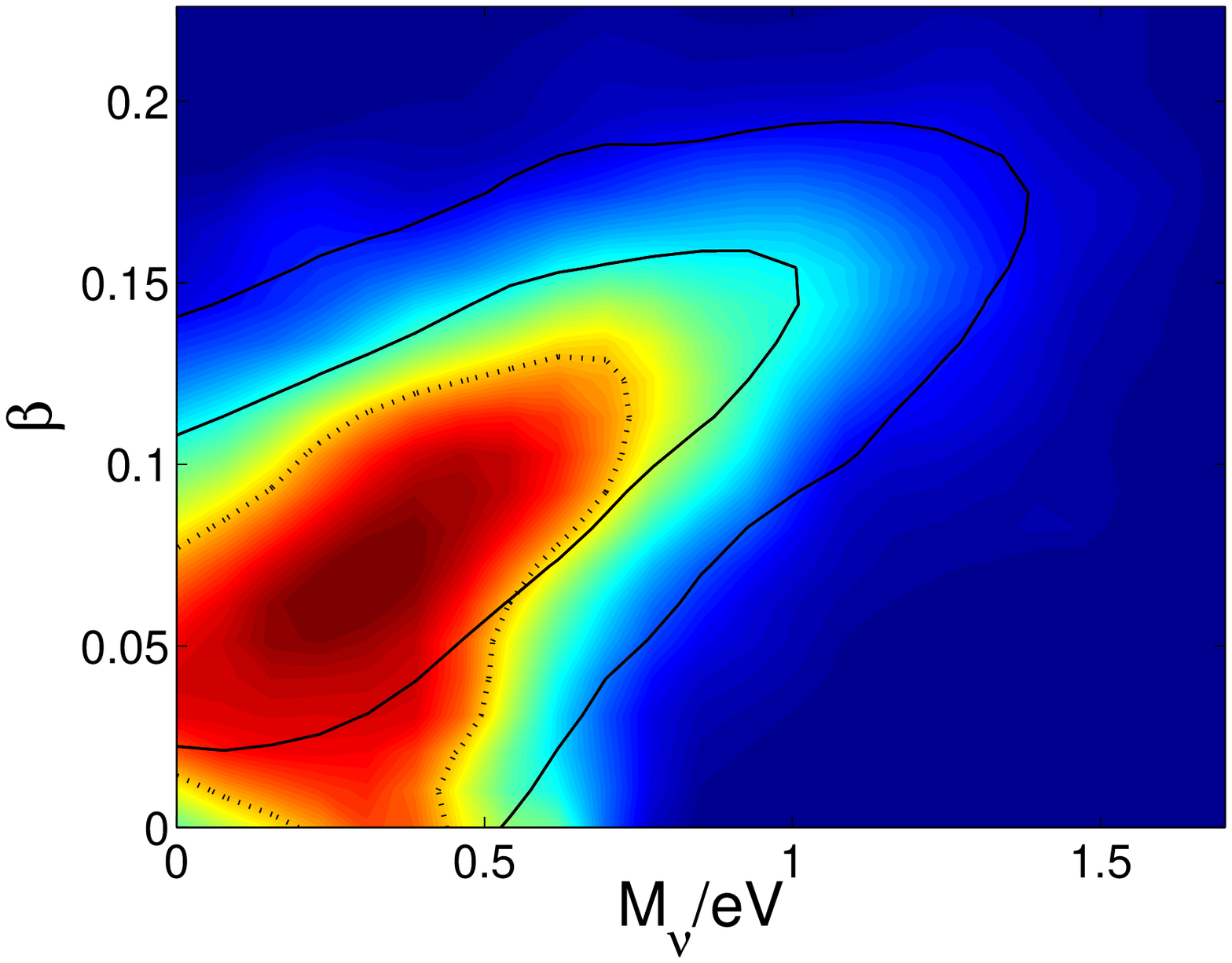}
\includegraphics[height=6.5cm,width=7.cm,angle=0]{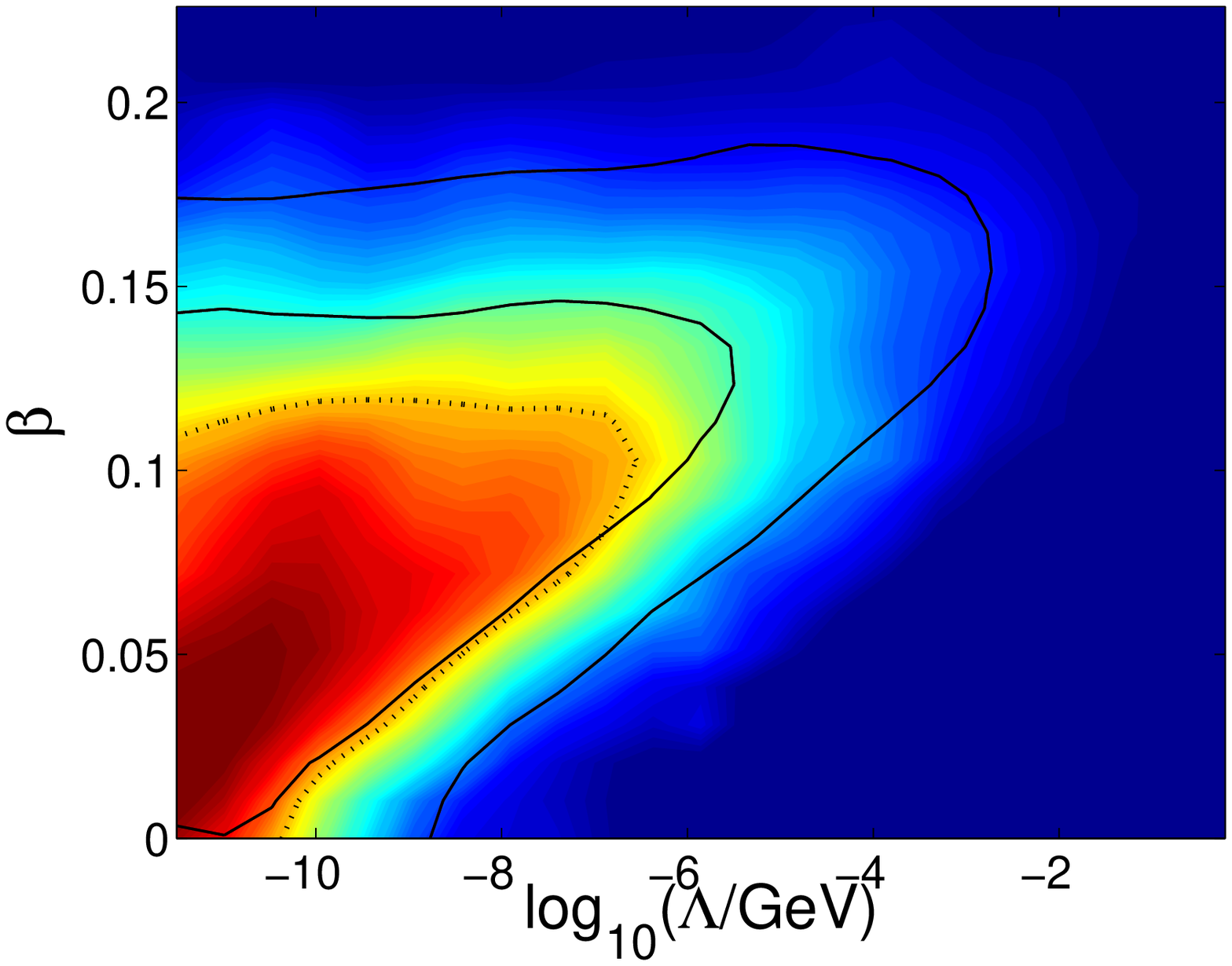}
\caption{As previous Figure, for a RP potential.}
\label{2DR}
\end{figure}

The prescription that the 2--component dark sector interacts with
baryons or radiation only through gravitaion reads $
T^{(c)~\mu}_{~~~~\nu;\mu} + T^{(de)~\mu}_{~~~\, ~~\nu;\mu} = 0 $
~(here $T^{(c,de)}_{\mu\nu}$ are the stress--energy tensors for CDM
and DE, let their traces read $T^{(c,de)}$). Only if we make the
further assumption that the 2 components are separate, it shall be $C
\equiv 0$ in the relations
\begin{equation}
\label{Tcouple}
T^{(de)~\mu}_{~~~\, ~~\nu;\mu} = +C T^{(c)} \phi_{,\nu}~,~~
T^{(c)~\mu}_{~~~~\nu;\mu} =- C T^{(c)} \phi_{,\nu}~,
\end{equation}
obtained by passing from the Jordan frame, where a Brans--Dicke frame
cosmology holds, to the Einstein frame \cite{coupling,brans}.  These
equations also show why DE cannot couple to any component with
vanishing stress--energy tensor trace, {\it e.g.} to radiation.

Besides of $C$, we shall also use the dimensionless coupling parameter
$ \beta = (3/16\pi)^{1/2} m_p C~.  $ If it is $C = 1/m_p$, therefore,
$\beta = 0.2443$; values $\beta \sim $$\cal O$$(0.1)$ mean then $C
\simeq 1/2m_p~$.

In this context let us then remind the interesting option of
considering DE coupled to $\nu$'s \cite{mavans}; the trace $T^{(\nu)}$
becomes significant only when their kinetic energy is redshifted to
values $\sim m_\nu$.

Another option considered in the literature is that the {\it r.h.s.}'s
of eqs.~(\ref{Tcouple}) are replaced by $ \pm \bar C T^{(de)}
\phi_{,\nu}$, where $\bar C$ sign is opposite to $C$ \cite{spagnoletti}.
This option causes a bootstrap effect, as the energy transfer to DE is
boosted as soon as its dilution is attenuated. In analogy with the
coupling to $\nu$'s, this approach therefore associates the
peculiarity of our epoch (the {\it coincidence}) with another {\it
accepted} peculiarity. Furthermore, at variance from our approach,
this option does not follow from any conformal transformation of
Brans--Dicke formulation.

It should be however noticed that the correlation between $M_\nu$ (the
trace of the 3-dimensional neutrino mass matrix) and $C$ seems no
longer to persist within the frame of this approach if new datasets
are considered \cite{spagnoletti2}.

\section{Procedure and Results}

The results shown in this paper are based on our generalization of the
public program CAMB \cite{camb}, enabling it to study cDE models.
Likelihood distributions are then worked out by using CosmoMC
\cite{lewis:2002}.

In Figure \ref{fig1a} we then illustrate the compensating effects
between coupling and $\nu$ mass, by considering various cosmologies.
A 00--model is then a cosmology without coupling and $\nu$ mass. In
turn, the 01-- and 10--models are cosmologies with coupling (and no
$\nu $ mass) and $\nu $ mass (and 0--coupling). In the 11--model,
finally, both $\nu $ mass and coupling are included.

The {\it l.h.s.} plot shows an example of transfer functions for 00--,
01--, 10-- and 11--models.  The sum of $\nu$ masses ($M_\nu$) is tuned
so that the 00-- and 11-- $T(k) $'s nearly coincide. At the {\it
r.h.s.}, we then see a similar comparison with $C_l$ anisotropy
spectral data. More details are provided in the caption.

More details on the fitting procedure are provided elsewhere
\cite{kristiansen09}. The fitting procedure returns parameter values 
mostly in the same range as for dDE or $\Lambda$CDM cosmologies. The
significant parameters in our approach are however the energy scale in
potentials ($\Lambda$), the coupling intensity ($\beta$), and the sum
of $\nu $ masses ($M_\nu$). In Figures \ref{mnuall} and \ref{Rmnuall}
we provide one--dimensional likelihood distributions on these
parameters for SUGRA and RP cosmologies.
\begin{figure}
\includegraphics[height=6.cm,width=8.2truecm,angle=0]{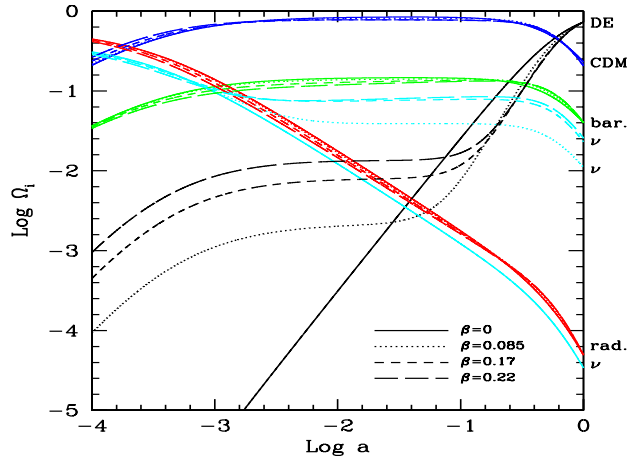}
\caption{Evolution of density parameters in a SUGRA model with
coupling and $\nu$ mass. Colors refer to different components, as
specified in the frame; lines to the different models: $M_\nu=0,
~\beta=0$ (continuous line); $M_\nu=0.5~$eV, $\beta = 0.085$ (dotted);
$M_\nu=1.1~$eV, $\beta = 0.17$ (short dashed); $M_\nu=1.2~$eV, $\beta
= 0.22$ (long dashed). }
\label{RPom}
\end{figure}
\begin{figure}
\includegraphics[width=11.cm,angle=0]{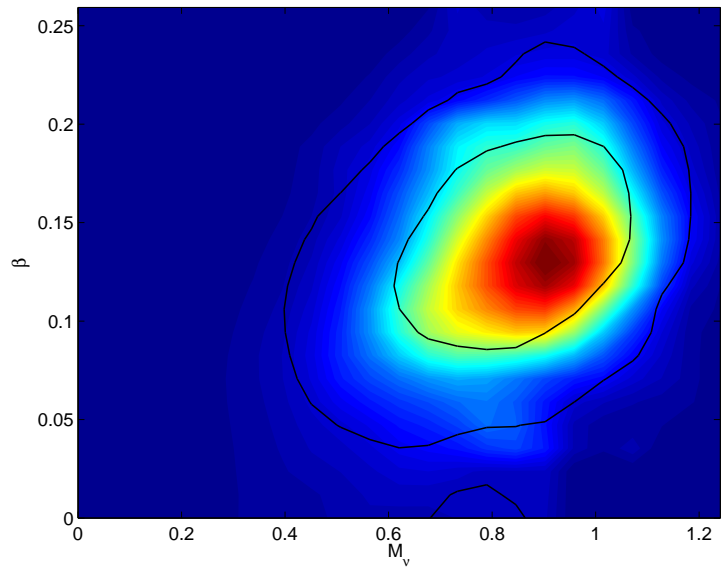}
\caption{Detection of CDM--DE coupling, following an hypothetical
determination of an electron neutrino mass $m_\nu \simeq 0.3~$eV, by
the experiment KATRIN, yielding $M_\nu \simeq 0.9~$eV.  Here we show
the likelihood distribution, with such a prior on $M_\nu$. Colors and
lines convey the same indications as in Figures 4 and 5.  }
\label{katrin}
\end{figure}

A basic information is however the correlation between likelihood
distributions. These are shown in Figures \ref{2DS} and \ref{2DR}
again for SUGRA and RP cosmologies, respectively.

These Figures, as well as one--dimensional plots, clearly exhibit
maxima, both for average and marginalized likelihood, for
significantly non--zero coupling and $\nu$--masses. Although their
statististical significance is not enough to indicate any
``detection'' level, the indication is impressive. Furthermore, a
stronger signal, with the present observational sensitivity, would be
impossible.

Let us however outline that this work aimed at finding how far one
could go from $\Lambda$CDM, adding non--zero coupling and $\nu$--mass,
without facing a likelihood degrade. It came then as an unexpected
bonus that likelihood does not peak on the 0--0 option.

\section{Conclusions}

The allowed $\beta$ values open the possibility of a critically
modified DE behavior. Figure \ref{RPom} shows the scale dependence of
the cosmic components for various $\beta$--$M_\nu$ pairs.

For $\beta$ values comprised between 0.1 and 0.22, {\it i.e.} well
within 2--$\sigma$'s from the best--fit model, just as the 0--0
option, we have a long plateau in the energy density of DE, going from
$z \sim 10$ to above $z = 1000$. The devline of $\rho_{de}$ at greater
$z$ is then mostly due to a parallel decline of the density $\rho_c$
of CDM: when radiation becomes dominant, both DCM and DE are
negligible.

The ratio $\rho_c/\rho_{de}$ in the plateau is $\cal O$$(100)$.  At
lower $z$ it becomes gradually smaller because of the gradual
contribution of the potential term to $\rho_{de}$. It should be
however outlined that this behavior is not {\it ad--hoc}, but is
incribed in the tracker solutions for the potentials selected.

Let us finally outline that the $M_\nu$ values allowed by $\beta \sim
0.1$--0.2 approach the $\nu$--mass detection area in the forthcoming
experiment KATRIN \cite{katrin}, based on tritium decay.  

Accordingly, should particle data lead to an external prior on
$M_\nu$, the strong degeneracy between the coupling parameter $\beta$
and the neutrino mass $M_\nu$ is broken, and new insight into the
nature of DE is gained \cite{kristiansen09}. In Figure \ref{katrin} we
show how a neutrino mass determination symultaneously implies an
almost model independent CDM--DE coupling detection.

This would be a revival of mixed DM models \cite{mix}, in the form of
Mildly Mixed Coupled (MMC) cosmologies.


\begin{theacknowledgments}

\end{theacknowledgments}


\newcommand{\Nature}{{\it Nature\/} }
\newcommand{\ApJ}{{\it Astrophys. J.\/} }
\newcommand{\ApJS}{{\it Astrophys. J. Suppl.\/} }
\newcommand{\MNRAS}{{\it Mon. Not. R. Astron. Soc.\/} }
\newcommand{\PhRv}{{\it Phys. Rev.\/} }
\newcommand{\PhL}{{\it Phys. Lett.\/} }
\newcommand{\JCAP}{{\it J. Cosmol. Astropart. Phys.\/} }
\newcommand{\AeA}{{\it Astronom. Astrophys.\/} }
\newcommand{\etall}{{\it et al.\/} }
\newcommand{\arXiv}{{\it Preprint\/} }

\end{document}